# ILP Aware Scheduling on Multithreaded Multi-core Processors

Murthy Durbhakula

Indian Institute of Technology Hyderabad, India

cs15resch11013@iith.ac.in, murthy.durbhakula@gmail.com

Abstract: Multithreaded Multi-core processors are prevalent today and are used for solving some of the important problems in computing. Resource imbalance can negatively impact overall performance in such processors. Hence balanced resource utilization is important in such processors. Particularly, it is important to maximize utilization of available instruction-level-parallelism (ILP). In this paper I present an ILP aware operating system (OS) scheduling algorithm for Multithreaded Multi-core processors. By keeping track of available ILP in each thread and by balancing it with available ILP resources in the system the OS will come up with a new schedule of threads for the next quantum. This new schedule will potentially result in a more balanced resource-utilization and improve performance for the next quantum. This work can be extended by doing a detailed quantitative evaluation.

## 1 Introduction

Multithreaded Multi-core processors are prevalent today and are used for solving some of the important problems in computing. For instance, Intel Xeon server processor [1] is a multi-threaded multi-core processor. Similarly IBM Power [2] and Oracle Niagara [3] class of processors are also multi-threaded multi-core processors. Balanced resource utilization is an important design aspect in such processors. Scheduling multiple low or high ILP workloads on the same multi-threaded processor will result in either low resource utilization when low ILP threads are scheduled together or stalls due to resource conflicts when high ILP threads are scheduled together. In this paper I present an Operating System (OS) scheduling policy which takes per-thread ILP into account while scheduling threads on multithreaded multi-core processors. By keeping track of ILP for each thread in every scheduling quantum the OS will determine where to schedule threads for the next quantum thereby resulting in a balanced utilization of available ILP and an improvement in overall performance. Here I assume there is no data sharing among threads. The rest of the paper is organized as follows: Section 2 presents ILP aware OS scheduling optimization. Section 3 describes related work and Section 4 presents conclusions.

## 2 OS Scheduling Optimization

We need dedicated hardware counters per thread that keeps track of average ILP per thread. This is the only hardware support we need for this optimization. Average ILP is calculated for a time window T which is less than the duration of scheduling quantum. The exact value of T needs to be determined empirically. This is to capture phase based behavior and not be biased by cumulative/historical behavior over large periods of time. For the next scheduling quantum the operating system will use these counters and try to schedule such a way that the cumulative ILP of threads are not more than available ILP in each multi-threaded processor. Sometimes cumulative ILP of threads may exceed available ILP in the system. This can happen if in a given scheduling quantum we end up with many high ILP threads.

Below is the pseudo-code of the scheduling algorithm.

Input: Threads T0, T1, ...TN with counters ILP0,ILP1,...ILPN.. Present schedule S_present which has mapping of N threads to N slots. For the sake of simplicity we consider a single socket system with K multi-threaded processors that can support L threads per processor. K*L = N.

Output: New schedule S_next with new mapping of threads to nodes

begin

1. Create a hash table of size $O(CN)$ entries where C is a constant. Take all N threads and hash their ILP value into appropriate entry in the hash table. Say if maximum ILP value is 4 and there are 128 threads in total and value of C is 4 then we create a hash table of size 512 entries and divide ILP value of 4 into 512 parts where each hash table entry indicates a step size of 4/512 = 0.0078125. So a thread with ILP value between 0 and 0.0078125 will sit in entry 0 of hash table. It is possible that multiple threads can hash to same entry of hash table creating a collision. In such cases those threads are sorted by their ILP value and are assigned to same hash table entry. In the worst case all threads may hash into same entry thus needing us to sort all N values. Hence worst case complexity of this step is $O(N \log N)$ and best case complexity is $O(N)$ as we need to walk through all N threads.

2. Walk through the hash table in descending order. Assign each of the top K ILP threads to K multi-threaded processors P0, P1,...PK. Assign next top K ILP threads in reverse order to P0, P1, ...PK. This is to balance the ILP allocation to K processors. So highest ILP thread in the second set of K ILP threads will go to PK instead of P0. For the third set of next highest K ILP threads we alternate back to original method of assigning highest ILP thread to P0. This way we do alternating assignments until we complete assigning all threads to all K processors.. This can be done in linear time as the number of entries in the hash table are $O(CN)$ where C is a constant, hence essentially $O(N)$.

end

Overall complexity: Overall complexity is dominated by step 1. Worst case complexity of step1 is $O(N \log N)$ and best case complexity is $O(N)$.

## 3 Related work

There are various alternative hardware and software solutions to maximize resource utilization in multi-threaded processors.

The retire unit can track which threads are retiring instructions close to maximum allowed limit per cycle and used that as a feedback mechanism to the thread arbiter. The thread arbiter can then prioritize picking those threads which are actively retiring instruction with least stalls. There should also be starvation avoidance mechanisms in-place to pick threads with low ILP from time-to-time.

There are various ILP enhancing compiler optimizations [4, 5]. Our work is orthogonal in the sense that it can also be applied to such compiler optimized workloads. Compiler optimizations have an advantage as they are software solutions and hence flexible. However in order for existing legacy applications to take advantage of this solution they need to be recompiled, which can be a drawback particularly if source code is not available.

## 4 Conclusions

Multithreaded Multi-core processors are prevalent today and are used for solving some of the important problems in computing. Resource imbalance can negatively impact overall performance in such processors. Hence balanced resource utilization is important in such processors. In this paper I have presented an OS scheduling optimization that keeps track of available ILP in every thread and uses that information to optimally schedule threads for the next scheduling quanta for balanced resource utilization as well as improvement of overall performance.